\documentstyle[12pt,axodraw]{article}
\def\e{{\rm e}}
\newcommand{\be}{\begin{equation}}
\newcommand{\ee}{\end{equation}}
\newcommand{\bea}{\begin{eqnarray}}
\newcommand{\eea}{\end{eqnarray}}
\newcommand{\al}{\alpha}

\newcommand{\gm}{\gamma}
\newcommand{\Gm}{\Gamma}
\newcommand{\dl}{\delta}

\newcommand{\eps}{\epsilon}
\newcommand{\ep}{\epsilon}

\newcommand{\pa}{\partial}
\newcommand{\dd}{\mbox{d}}

\newcommand{\nn}{\nonumber}
\newcommand{\uk}{\underline{k}}

\newcommand{\ual}{\underline{\alpha}}

\def\pppp{{\bf 4}^+}

\def\pppppp{{\bf 6}^+}

\def\mmmmm{{\bf 5}^-}
\def\pnine{{\bf 9}^+}
\newcommand{\Li}[2]{{\mbox{Li}}_{#1}\left(#2\right)}
\newcommand{\z}{&\hspace*{-8pt}}
\renewcommand{\eps}{\varepsilon}

\begin{document}
\parindent=1.5pc

\begin{titlepage}
\begin{flushright}
\end{flushright}
\vskip 1.8cm
\begin{center}
 \boldmath
{\Large\bf
Analytical Results for Dimensionally Regularized \\
Massless On-shell Double Boxes with Arbitrary Indices and
Numerators} \unboldmath
\vskip 1.2cm
{\sc V.A. Smirnov\footnote{E-mail:
smirnov@theory.npi.msu.su}}
\vskip .3cm
{\em Nuclear Physics Institute, Moscow State University, \\
119889 Moscow, Russia}
\vskip .7cm
{\sc O.L. Veretin\footnote{E-mail: veretin@ifh.desy.de  }}
\vskip .3cm
{\em DESY, 15738 Zeuthen, Germany }
\vskip 1.0cm
\end{center}

\begin{abstract}
\noindent
We present an algorithm for the analytical evaluation of
dimensionally regularized massless on-shell double box Feynman
diagrams with arbitrary polynomials in numerators and general integer
powers of propagators. Recurrence relations following from
integration by parts are solved explicitly and any given double box
diagram is expressed as a linear
combination of two master double boxes and a family of simpler diagrams.
The first master double box
corresponds to all powers of the propagators equal to one and
no numerators, and the second master double box differs from the first one by
the second power of the middle propagator.
By use of differential relations, the second master
double box is expressed through the first one up to a similar
linear combination of simpler double boxes so that the analytical
evaluation of the first master double box provides explicit analytical
results, in terms
of polylogarithms $\Li{a}{ -t/s }$, up to $a=4$, and generalized
polylogarithms $S_{a,b}(-t/s)$, with $a=1,2$ and $b=2$, dependent on
the Mandelstam variables $s$ and $t$,
for an arbitrary diagram under consideration.
\end{abstract}

\vfill

\end{titlepage}

\section{Introduction}

The massless double box diagram shown in Fig.~1 with general numerators
and integer powers of propagators is relevant to
many important physical processes.
The purpose of the present paper is
to present an algorithm for the analytical evaluation of
the general massless on-shell (i.e. for $p_i^2=0,\;
i=1,2,3,4$) double box Feynman diagram
in the framework of dimensional regularization \cite{dimreg},
with the space-time dimension $d=4-2\ep$ as a regularization parameter.
%
\vspace{0.5cm}
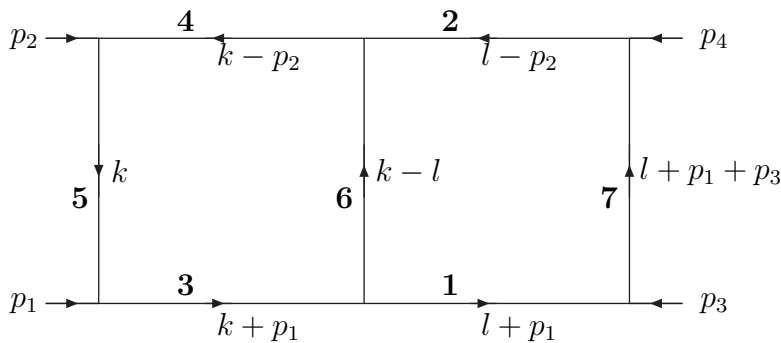
\begin {figure} [htbp]
\begin{picture}(400,100)
\ArrowLine(80,0)(100,0)
\ArrowLine(80,100)(100,100)
\ArrowLine(320,0)(300,0)
\ArrowLine(320,100)(300,100)
\Line(80,0)(320,0)
\Line(80,100)(320,100)
\Line(100,0)(100,100)
\Line(200,0)(200,100)
\Line(300,0)(300,100)
\ArrowLine(150,100)(140,100)
\ArrowLine(250,100)(240,100)
\ArrowLine(140,0)(150,0)
\ArrowLine(240,0)(250,0)
\ArrowLine(100,60)(100,40)
\ArrowLine(300,40)(300,60)
\ArrowLine(200,40)(200,60)
\Text(105,50)[l]{$ k$}
\Text(205,50)[l]{$ k- l$}
\Text(305,50)[l]{$ l+p_1+p_3$}
\Text(145,-9)[l]{$ k+p_1$}
\Text(245,-9)[l]{$ l+p_1$}
\Text(145,93)[l]{$ k-p_2$}
\Text(245,93)[l]{$ l-p_2$}
\Text(230,107)[l]{\bf 2}
\Text(130,107)[l]{\bf 4}
\Text(230,7)[l]{\bf 1}
\Text(130,7)[l]{\bf 3}
\Text(90,40)[l]{\bf 5}
\Text(190,40)[l]{\bf 6}
\Text(290,40)[l]{\bf 7}
\Text(67,0)[l]{$p_1$}
\Text(67,100)[l]{$p_2$}
\Text(328,0)[l]{$p_3$}
\Text(328,100)[l]{$p_4$}
\end{picture}
\vspace*{2mm}\\
\caption{Planar double box diagram.}
\end{figure}
\vspace*{0.0cm}\\
The dimensionally regularized on-shell master double box Feynman integral
(i.e. with all powers of propagators equal to one and no numerators)
has been analytically evaluated in \cite{K1} in terms
of polylogarithms $\Li{a}{ -t/s }$, up to $a=4$, and generalized
polylogarithms $S_{a,b}(-t/s)$, with $a=1,2$ and $b=2$, dependent on
the Mandelstam variables $s=(p_1+p_2)^2$ and $t=(p_1+p_3)^2$.
In \cite{V1}, recurrence relations within the method of
integration by parts (IBP) \cite{IBP} were explicitly solved and any given double
box diagram was expressed as a linear
combination of the master double box mentioned above, the
second master double box which differs from the first one by
the second power of the middle propagator,
a family of double boxes with two contracted lines considered in shifted
dimension, and vertex diagrams.

In the present paper, we complete this algorithm by evaluating the
second master double box and presenting crucial checks
of our results by use of asymptotic expansions in the limits $t/s \to 0$ and
$s/t \to 0$. In the next section, we present definitions of the
double box diagrams through integrals in loop momenta and
$\alpha$-parameters. Then we describe recurrence relations and their
solutions in terms of the two master double boxes
and a collection of simpler diagrams, in accordance with~\cite{V1}.
In Section~3, we describe how the double boxes with
two contracted lines are analytically evaluated in expansion in $\ep$
up to a desired order. In Section~4, we present the analytical result
of ref.~\cite{K1}
for the first master double box and then use
differential relations in order to express the second master
double box through the first one up to a similar
linear combination of simpler double boxes. This provides an explicit
analytical result for the second master double box, in terms of the
same class of functions as for the first one.
In Section~5, we use the general strategy of regions for expanding the double
boxes in the limits $t/s \to 0$ and $s/t \to 0$. We present analytical
algorithms for the evaluation of the hard-hard and collinear-collinear
contributions to the expansion.
We evaluate the LO and NLO (respectively order $1/(s^2 t)$ and $1/(s^3)$)
of the expansion of the first double box
in the limit $t/s \to 0$ and find an agreement with our explicit
result. In conclusion we discuss our results.

\section{Recurrence relations and their solution}

The general massless on-shell double box Feynman integral
in $d$-dimension can be written as
\bea
K^{(d)}(P,a_1,\ldots,a_7;s,t) =
\int\int \frac{\dd^dk\, \dd^dl }{(\pi^{d/2})^2} \,\frac{1}{(l^2+2 p_1 l)^{a_1} (l^2-2 p_2 l)^{a_2} }
\nn \\ \times
\frac{P(p_1,p_2,p_3,k,l)}{(k^2+2 p_1 k)^{a_3}
(k^2-2 p_2 k)^{a_4} (k^2)^{a_5} ((k-l)^2)^{a_6}
((l-p_1-p_3)^2)^{a_7} } \, ,
\label{2box}
\eea
where $P$ is a polynomial, $a_i$ integers,
$k$ and $l$ are respectively loop momenta of the left and the right box.
Usual prescriptions, $k^2=k^2+i 0, \; -s=-s-i 0$, etc. are implied.

The $\alpha$-representation of the double box with $P\equiv 1$
is straightforwardly obtained (we omit in the following
the polynom $P$ and consider only scalar integrals):
\bea
K^{(d)}(a_1,\ldots,a_7;s,t) =
\frac{(-1)^{a_1+\ldots+a_7} i^{a_1+\ldots+a_7+ 2-d}}{\prod_i \Gm(a_i)}
\nn \\ \times
\int_0^\infty \dd\al_1 \ldots\int_0^\infty\dd\al_7
\prod_i  \al_i^{a_i-1}  D^{-d/2}
\exp \left[i\frac{A}{D}s + i \frac{\al_5\al_6\al_7}{D} t \right]
 \; ,
\label{KGenAlpha}
\eea
where
\bea
D&=&(\al_1+\al_2+\al_7) (\al_3+\al_4+\al_5)
+\al_6 (\al_1+\al_2+\al_3+\al_4+\al_5+\al_7) \;,
\label{Dform}
\\
A&=& \al_1\al_2 (\al_3+\al_4+\al_5) + \al_3\al_4(\al_1+\al_2+\al_7)
+\al_6 (\al_1+\al_3)( \al_2+\al_4)
\label{Aform} \; .
\eea

  To deal with Feynman integrals with numerators we use the fact
that any polynomial in the numerators
of the propagators can be represented as a differential operator with respect to
some auxiliary parameters (see, e.g., \cite{Breitenlohner}) acting
on a scalar diagram. An outcome of this procedure is
that any tensor integral is expressed in terms of scalar
integrals but in different (shifted) space-time dimensions
and with shifted indices of lines (see a detailed discussion in \cite{OTdim}).
This step is straightforward and formulae from \cite{OTdim} can be easily
programmed on computer.

  A more difficult part of the program is to express the so obtained
double box integrals (in different dimensions and with all possible
sets of indices) in terms of some master integrals and a family of simpler
boundary integrals. It turns out that, in our case, there arise
only two master integrals
$K^{(d)}_{(1)}=K^{(d)}(1,1,1,1,1,1,1;s,t),
\,K^{(d)}_{(2)}=K^{(d)}(1,1,1,1,1,2,1;s,t)$, and
the boundary integrals are either vertex integrals that are
evaluated in gamma functions and integrals with at least two reduced lines.

   Using the integration by parts method \cite{IBP} the following relation
can be obtained to reduce the index $a_1$ to unity:
\begin{eqnarray}\label{planarRR1}
\z\z s a_1{\bf 1^+} =
    a_7{\bf 7^+2^-}+ a_6{\bf 6^+}({\bf 2^-}-{\bf 4^-})
  + a_1{\bf 1^+2^-}-(d-2 a_2- a_1- a_7- a_6)\,.
\end{eqnarray}
Hereafter we use the standard notation: $\bf j^{\pm}$ is the
operator increasing/decreasing the index on the $j$th line by
one unit, i.e. ${\bf j^{\pm}}K(\dots,a_j,\dots)=K(\dots,a_j\pm1,\dots)$.

  Three similar relations obtained by permutations of lines
can be used to reduce indices of lines 1,2,3 and 4 to one.
Next we can reduce indices of lines 5 and 7 with the help of the following
relations:
\begin{eqnarray}
\z\z (d-2-2 a_5- a_4- a_3) a_5{\bf 5^+} =
  (d-2-2 a_6- a_4- a_3) a_6{\bf 6^+}  \nonumber\\
\z\z\qquad
  +( a_5- a_6) a_4{\bf 4^+}
  +( a_5- a_6) a_3{\bf 3^+}
  + a_4 a_6{\bf 4^+6^+2^-}
  + a_3 a_6{\bf 3^+6^+1^-}\,, \\
\z\z\nonumber\\
\z\z (d-2-2 a_7- a_2- a_1) a_7{\bf 7^+} =
  (d-2-2 a_6- a_2- a_1) a_6{\bf 6^+}  \nonumber\\
\z\z\qquad
  +( a_7- a_6) a_2{\bf 2^+}
  +( a_7- a_6) a_1{\bf 1^+}
  + a_2 a_6{\bf 2^+6^+4^-}
  + a_1 a_6{\bf 1^+6^+3^-}.
\label{planarRR7}
\end{eqnarray}
Using the above recurrence relations
we can bring indices of lines 1,2,3,4,5,7 all to unity
so that only $a_6$ can be greater than one.

   Our relation to reduce the index of line 6 reads \cite{V1}
\begin{eqnarray}
\label{P6}
\z\z t(d-6-2 a_6)( a_6+1) a_6{\bf 6^{++}}  = \nonumber\\
\z\z\qquad
     -(d-5- a_6)\bigl[ 3d-14-2 a_6 + 2 a_6\frac{t}{s} \bigl]  a_6{\bf 6^+} \nonumber\\
\z\z\qquad
     + \frac2s (d-4- a_6)^2(d-5- a_6) \nonumber\\
\z\z\qquad
     +\Biggl\{
       ({\bf2^+}+{\bf7^+}) \Bigl[
                          -\frac2s (d-4- a_6)(d-5- a_6)
                          +2\frac{t}{s} a_6^2 {\bf6^+}
                            \Bigr]    \nonumber\\
\z\z\qquad
    -   \Bigl[ 2t( a_6+1) a_6{\bf6^{++}} + 2(d-4- a_6) a_6{\bf6^+}
              \Bigr] {\bf 3^+}
      \Biggr\} {\bf1^-}  \nonumber\\
\z\z\qquad
    + (d-6){\bf7^-}{\bf d^-}\,,
\end{eqnarray}
where $\bf d^-$ decreases the dimension of space-time by 2.
Note that the dimension can be effectively shifted
only by an even integer number (see detailed discussion in \cite{OTdim}).
We stress that this formula is valid only if the indices of
lines 1,2,3,4,5,7
were already reduced to unity. Note that in the left-hand side of (\ref{P6})
there is $\bf6^{++}$, rather than $\bf6^+$. This means that
the index of line~6 cannot be always reduced to one but generally to one
or two. 
One can also get rid of $\bf d^-$ in the above formulae
replacing it by 
$({\bf1^+}+{\bf2^+}+{\bf6^+})({\bf3^+}+{\bf4^+}+{\bf5^+})
+{\bf6^+}({\bf1^+}+{\bf2^+})$.

  Let us comment shortly on how (\ref{P6}) and other similar
relations can be derived. One can start from the integral with
numerator $2kp_2$. Since $2kp_2=(k-p_2)^2-k^2$ we can eliminate
this numerator by canceling lines 4 or 5 (see Fig.~1), and the
resulting integrals are simple. On the other hand one can
use the machinery mentioned after (\ref{Aform}) 
(see \cite{Breitenlohner,OTdim}) which expresses tensor integrals
as differential operator acting on scalar integral in some
dimension $d+2n$. It is more convinient in our case to start
from the dimension $d-2$. Then we have
\begin{equation}
(2kp_2){\bf d^-} =
    -s(\partial_1\partial_6+\partial_2\partial_3
         +\partial_3\partial_6+\partial_3\partial_7 )
         +t (\partial_6\partial_7 ),
\end{equation}
where $\partial_j=\partial/\partial m^2_j$ takes the derivative
with respect to the square of the mass on $j$th line. 
(After differentiaton
all masses are put to zero). In the right-hand side of the above formula,
there are scalar integrals with increased indices. Therefore we
can apply reduction formulae (\ref{planarRR1})--(\ref{planarRR7}).
The resulting relation will involve terms like $\bf6^{++}$ and, after
some transformation, one can come to (\ref{P6}).

  To complete the reduction procedure we should bring the two
master integrals (which can appear in shifted dimensions) to
the integrals in the generic $d=4-2\varepsilon$ dimension.
Thus we need also a relation that reduces the dimension of
the space-time. This can be obtained by inverting the identity 
\begin{equation}
  K^{(d-2)}(a_1,\dots) = D(\partial) K^{(d)}(a_1,\dots),
\label{Drelation}
\end{equation}
where $K^{(d)}$ is defined in (\ref{2box}),
$D(\alpha)$ is given by (\ref{Dform}), and
$\partial$ denotes a family of differential operators acting on auxiliary
masses of the lines: $\partial_i = \partial/\partial m^2_i$.
(After the differentiation, all these masses are put to zero.)

(\ref{Drelation}) is valid for arbitrary indicies $a_i$ and can be
derived from the $\alpha$-representation of the Feynman integral
(see e.g. \cite{OTdim}). This relation however increases
the dimension by 2 units. To find a relation decreasing dimension
we have to compute operator $D^{-1}$ which is inverse to $D$.
These can be done with the help of already listed above reduction
formulae.  Instead of giving general form of inverse to (\ref{Drelation})
it is enough to give it for the master integrals.
Since we have two master integrals
there are two relations \cite{V1}
\begin{eqnarray}
K_{(1)}^{(d)} \z=\z \frac{1}{\Delta} \Bigl[
        + a_{22} ( K_{(1)}^{(d-2)} - f_1^{(d)} K^{(d)}_{(1)})
           - a_{12} (K_{(2)}^{(d-2)} - f_2^{(d)} K^{(d)}_{(1)}) \Bigr],\\
K_{(2)}^{(d)} \z=\z \frac{1}{\Delta} \Bigl[
        -a_{21} ( K_{(1)}^{(d-2)} - f_1^{(d)} K^{(d)}_{(1)})
           + a_{11} (K_{(2)}^{(d-2)} - f_2^{(d)} K^{(d)}_{(1)} ) \Bigr],
\end{eqnarray}
where operators $f^{(d)}_j$ are given by
\begin{eqnarray}
f_1^{(d)}\z\z =   \nonumber\\
\z\z + \Biggl\{
       \frac2s({\bf2^+3^+}+{\bf2^+4^+}+{\bf2^+6^+}+{\bf4^+6^+}+{\bf4^+7^+}
                 +{\bf3^+7^+}) \nonumber\\
\z\z
        + \frac4s({\bf2^+5^+}+{\bf5^+6^+}+{\bf5^+7^+})
  -\frac{2}{s^2t}(d-5)(3s+2t)({\bf2^+}+{\bf7^+})  \nonumber\\
\z\z
  + \frac{2}{d-6}{\bf3^+6^+7^+}
       -\frac{2}{st(d-6)} \Bigl( 3s(d-5)+t(3d-14) \Bigr) {\bf3^+6^+}
      \Biggr\} {\bf 1^-} \nonumber\\
\z\z
   +\frac3t {\bf7^-d^-} \, ,
\end{eqnarray}

\begin{eqnarray}
f_2^{(d)}\z\z =   \nonumber\\
\z\z + \Biggl\{
       \frac2s({\bf2^+3^+}+{\bf2^+4^+}+{\bf3^+7^+}+{\bf4^+7^+}){\bf6^+}
        + \frac4s({\bf2^+}+{\bf4^+}){\bf6^{++}} \nonumber\\
\z\z
    +\frac{2(2d-13)}{s(d-6)}({\bf2^+}+{\bf7^+}+2\cdot{\bf6^+}){\bf5^+6^+}
              \nonumber\\
\z\z
   - \frac{2(d-5)(d-7)}{s^2t(d-6)(d-8)}\Bigl( s(3d-20)+2t(d-6) \Bigr)
        ({\bf2^+}+{\bf7^+}){\bf6^+}   \nonumber\\
\z\z
  + \frac{2(d-5)(d-7)}{s^2t^2(d-8)} \Bigl( 3s(3d-20)+4t(2d-13) \Bigr)
           ({\bf2^+}+{\bf7^+} + \frac{s}{d-6}{\bf3^+6^+})  \nonumber\\
\z\z
  + \frac{4}{d-6} ( \frac1s + {\bf3^+} ) {\bf7^+6^{++}}
   +\frac{4}{d-8}\Bigl( \frac{5d-34}{s} + \frac{(3d-20)(2d-13)}{t(d-6)} \Bigr)
          {\bf3^+6^{++}}
      \Biggr\} {\bf1^-}  \nonumber\\
\z\z
  + \Biggr\{
    \frac{3d-20}{t(d-6)}{\bf6^+} - \frac{d-7}{st^2(d-8)}\Bigl( 3s(3d-20)+4t(2d-13) \Bigl)
     \Biggr\} {\bf7^-d^-} \, ,
\end{eqnarray}

\begin{eqnarray}
a_{11} \z=\z \frac{2}{s^2t}(d-5)^2(3s+2t),  \\
a_{12} \z=\z -\frac{2}{s}(4d-21) -\frac3t (3d-16),\\
a_{21} \z=\z -\frac{(d-5)^2(d-7)}{st(d-8)}
     \Bigl( \frac{8(2d-13)}{s} + \frac{6(3d-20)}{t} \Bigr),\\
a_{22} \z=\z \frac{d-7}{s^2t^2(d-8)}
     \Bigl( 3s^2(3d-16)(3d-20) + 6st(5d^2-59d+172) \nonumber\\
        \z\z + 4t^2(d-5)(d-6) \Bigr),\\
\Delta \z=\z
     \frac{16(s+t)(d-5)^3(d-6)(d-7)}{s^4t(d-8)} \, .
\label{planarRRlast}
\end{eqnarray}

  Formulae (\ref{planarRR1})--(\ref{planarRRlast}) solve the problem
of the reduction of any planar double box to two master integrals
$K_{(1)},\,K_{(2)}$
and a set of simpler integrals with reduced lines.
{}From now on we omit the superscript $(d)$ because we shall deal
with integrals in $d=4-2\ep$ (non-shifted) dimensions.

  We have checked the reduction scheme 
(\ref{planarRR1})--(\ref{planarRRlast})
by expanding the integrand in two regimes, $s/t\to 0$ and $t/s\to 0$,
and evaluating the resulting one-scale integrals by the method
described in Sect.~5. 

\section{Double boxes with two contracted lines}

In the framework of the reduction procedure presented in the previous section,
the boundary values for general double boxes are either two master double
boxes, or vertex diagrams, i.e. at $a_5=0$ or $a_7=0$, or double boxes
with two contracted lines. The latter can be of the following two types:
$a_1=a_4=0$ (or the symmetrical variant $a_2=a_3=0$) shown in Fig.2,
or $a_3=a_4=0$ (or the symmetrical variant $a_1=a_2=0$) shown in Fig.3. Let us call them
respectively the box with a diagonal and the box with a one-loop insertion.
Note that they generally arise in a shifted dimension.

 We should mention that there are no integrals with
only one contracted line left. As far as one of the lines 1,2,3 or 4
is contracted one can still proceed with reduction formulae
from Sect. 2 or apply the standard ``rule of triangle'' \cite{IBP}
to reduce one more line. Thus the two master double boxes plus
boxes with two contracted line form the basis of integrals.
%
\vspace{0.5cm}
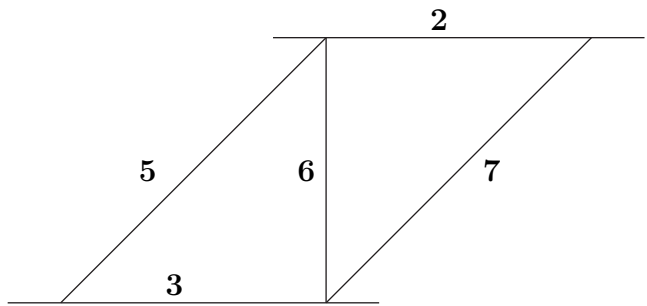
\begin {figure} [htbp]
\begin{picture}(400,100)
\Line(80,0)(220,0)
\Line(180,100)(320,100)
\Line(100,0)(200,100)
\Line(200,0)(200,100)
\Line(200,0)(300,100)
\Text(240,107)[l]{\bf 2}
\Text(140,7)[l]{\bf 3}
\Text(130,50)[l]{\bf 5}
\Text(190,50)[l]{\bf 6}
\Text(260,50)[l]{\bf 7}
\end{picture}
\vspace*{0.0cm}\\
\caption{A box with a diagonal.}
\end{figure}
\vspace*{0.0cm}\\
%
\vspace{0.5cm}
\begin {figure} [htbp]
\begin{picture}(400,100)
\Line(80,0)(220,0)
\Line(80,100)(220,100)
\Line(100,0)(100,100)
\CArc(300,50)(111.803,153.435,206.565)
\CArc(100,50)(111.803,-26.5651,26.5651)
\Text(130,107)[l]{\bf 4}
\Text(130,7)[l]{\bf 3}
\Text(90,50)[l]{\bf 5}
\Text(170,50)[l]{\bf 6}
\Text(225,50)[l]{\bf 7}
\end{picture}
\vspace*{0.0cm}\\
\caption{A box with a one-loop insertion.}
\end{figure}
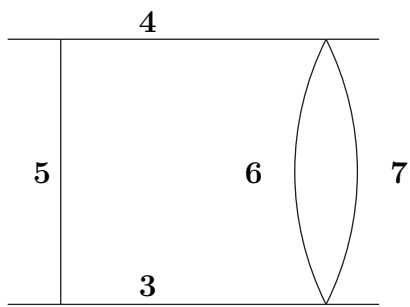
\vspace*{0.0cm}\\
%

  These cases without two lines are much simpler than the master double boxes.
Using $\alpha$-representation (\ref{KGenAlpha}) and representing one of the
functions involved into the Mellin--Barnes (MB) integral
\be
\frac{1}{(X+Y)^{\nu}} = \frac{1}{\Gm(\nu)}
\frac{1}{2\pi i}\int_{-i \infty}^{+i \infty} \dd w
\frac{Y^w}{X^{\nu+w}} \Gm(\nu+w) \Gm(-w) \; ,
\label{MB}
\ee
we obtain the following results:
\bea
K^{(d)}(0,a_2,a_3,0,a_5,a_6,a_7;d+n;s,t) & = &
\frac{i^2 (-1)^{a}}{\prod \Gm(a_i)
(-s)^{a -n-4+2\ep }}
\nn \\ && \hspace*{-71mm}
\times \frac{
\Gm(2 - a_3 - a_5 - \ep + n /2)
    \Gm(2 - a_6 - \ep + n /2)
    \Gm(2 - a_2 - a_7 - \ep + n /2)}{
\Gm(6 - a +3n/2-3\ep)
    \Gm(4 - a_3 - a_5 - a_6 +n-2\ep)
    \Gm(4 - a_2 - a_6 - a_7 +n-2\ep)}
\nn \\ && \hspace*{-71mm}
\times
\frac{1}{2\pi i}\int_{-i \infty}^{+i \infty} \dd z \,
 (t/s)^z  \Gm(a_5 + z) \Gm(a_7 + z) \Gm( a -4 -n+2\ep + z) \Gm(-z)
\nn \\ && \hspace*{-71mm}
\times \Gm(4 - a_2 - a_5 - a_6 - a_7 +n-2\ep - z)
    \Gm(4 - a_3 - a_5 - a_6 - a_7 +n-2\ep - z)
 \, ,
\label{BoxDiagMB}
\eea
and
\bea
K(a_1,a_2,0,0,a_5,a_6,a_7;d+n;s,t) &=&
\frac{i^2 (-1)^{a}}{\prod \Gm(a_i)
(-s)^{a -n-4+2\ep }}
\nn \\ && \hspace*{-70mm}
\times \frac{\Gm(2 - a_5 - \ep + n /2) \Gm(2 - a_6 - \ep + n /2) }{
\Gm(6 - a +3 n/2-3\ep) \Gm(4 - a_5 - a_6 +n-2\ep)}
\nn \\ && \hspace*{-70mm}
\times \frac{1}{2\pi i}\int_{-i \infty}^{+i \infty} \dd z \,
 (t/s)^z   \Gm(a_7 + z) \Gm(-4 + a -n+2\ep + z)
\nn \\ && \hspace*{-70mm}
\times \Gm(-2 + a_5 + a_6 + \ep - n /2 + z)
\Gm(4 - a_1 - a_5 - a_6 - a_7 +n-2\ep - z)
\nn \\ && \hspace*{-70mm}
\times
    \Gm(4 - a_2 - a_5 - a_6 - a_7 +n-2\ep - z) \Gm(-z)  \, ,
\label{BoxInsertMB}
\eea
where $a=\sum_i a_i$.
The contour of integration is chosen in the standard way:
the poles with the $\Gm(\ldots+z)$-dependence are to the left of the contour and
the poles with the $\Gm(\ldots-z)$-dependence are
to the right of it.

Then each of these boundary integrals with given values of integer
indices is decomposed into `singular' and `regular' parts.
For the diagonal crossed boxes, the singular part is written as
minus the sum of the residua of the integrand at the points $j-2 \ep$,
with $j=-{\rm{max}}\{a_2,a_3\}-a_5-a_6-a_7+4+n,\ldots,-1$,
plus the sum of the residua of the integrand at the points $j-2 \ep$
for $j=0,\ldots, 4+n-a_2-a_3-a_5-a_6-a_7$.
For the boxes with one-loop insertions, the singular part is written as
minus the sum of the residua of the integrand at the points $j-2 \ep$,
with $j=-{\rm{max}}\{a_1,a_2\}-a_5-a_6-a_7+4+n,\ldots,-1$,
plus the sum of the residua of the integrand at the points $j-2 \ep$
for $j=0,\ldots, 4+n-a_1-a_2-a_5-a_6-a_7$.

The regular parts are given by MB integrals where no gluing of poles
of gamma functions with $+z$ and $-z$ dependence arises. They can be
written as MB integrals for $-1<\mbox{Re}(z)<0$ with an integrand
expanded in a Laurent series in $\ep$ up to a desired order.
Then these integrals are straightforwardly evaluated by closing
the contour of integration to the right and taking residua at
the points $z=0,1,2,\ldots$. At this step, one can use a collection
of formulae for summing up series presented in ref.~\cite{Oleg}.
The evaluation of both the singular and the regular parts is easily
realized on computer.

Note that in the recurrence relations of the previous sections,
these boundary double boxes can arise with coefficients that
involve poles up to the second order in $\ep$ so that the expansion
up to $\ep^2$ is here necessary. However the `master' boxes with a diagonal 
or one-loop insertion enter with coefficients finite in $\ep$
so that it is sufficient to compute then, in expansion in $\ep$,
up to the finite part.
 Let us, for example, present an analytical result for the master box 
with a diagonal
in $d$ dimensions:
\be
K(0,1,1,0,1,1,1;d;s,t) =
\frac{\left(i \e^{-\gm_{\rm E}\ep} \right)^2 }{-s-t}
K_0(s,t,\ep) \, ,
\ee
where
\bea
K_0(s,t,\ep) &= &
- \left( \ln^2 (t/s) +\pi^2  \right) \frac{1}{2\ep^2}
\nn \\ && \hspace*{-27mm}
+ \left[
2 \Li{3}{ -t/s } -2\ln (t/s) \Li{2}{ -t/s }
-\left( \ln^2 (t/s) +\pi^2 \right) \ln(1+t/s)
\right. \nn \\ && \hspace*{-27mm} \left.
+ \frac{2}{3}\ln^3 (t/s) + \ln (-s) \ln^2 (t/s) + \pi^2 \ln (-t)
-2\zeta(3)
\right] \frac{1}{\ep}
\nn \\ && \hspace*{-27mm}
+ 4 \left(S_{2,2}(-t/s) - \ln (t/s) S_{1,2}(-t/s)  \right)
- 4 \Li{4}{ -t/s }
\nn \\ && \hspace*{-27mm}
+ 4 \left(\ln(1+t/s) -  \ln (-s)  \right) \Li{3}{ -t/s }
\nn \\ && \hspace*{-27mm}
+ 2\left( \ln^2 (t/s) +2 \ln (-s) \ln(t/s) -2\ln(t/s) \ln(1+t/s)
\right) \Li{2}{-t/s}
\nn \\ && \hspace*{-27mm}
+ 2\left(
\frac{2}{3}\ln^3 (t/s) + \ln (-s) \ln^2 (t/s) + \pi^2 \ln (-t)
-2\zeta(3)
\right) \ln(1+t/s)
\nn \\ && \hspace*{-27mm}
-\left( \ln^2 (t/s) +\pi^2  \right) \ln^2(1+t/s)
-\frac{1}{2} \ln^4 (t/s)-\frac{4}{3}\ln (-s) \ln^3 (t/s)
\nn \\ && \hspace*{-27mm}
-\left( \ln^2 (-s) + \frac{11}{12}\pi^2  \right) \ln^2(t/s)
-\pi^2 \ln^2(-s) -2\pi^2 \ln (-s) \ln(t/s)
\nn \\ && \hspace*{-27mm}
+ 4 \zeta(3)\ln(-t) -\frac{\pi^4}{20}
 \; .
\label{K0}
\eea
Here $\Li{a}{z}$ is the polylogarithm \cite{Lewin} and
\be
\label{Sab}
  S_{a,b}(z) = \frac{(-1)^{a+b-1}}{(a-1)! b!}
    \int_0^1 \frac{\ln^{a-1}(t)\ln^b(1-zt)}{t} \dd t \;
\ee
a generalized polylogarithm \cite{GenPolyLog}.
Using known formulae that relate polylogarithms and generalized polylogarithms
with arguments $z$ and $1/z$ \cite{Lewin,GenPolyLog} one can rewrite
this and similar results for the master double boxes in terms of
the same class of functions depending on the inverse ratio $s/t$.

  We do not give explicit result for the box with a 1-loop
insertion since it is certainly a simpler object. 
Indeed the 1-loop onsertion can be trivially integrated
and we have a 1-loop box where one of the indicies is equal to
$\varepsilon$. Such kind of integrals can be expressed 
using standard methods in terms
of the hypergeometric function ${}_2F_1$ (see e.g. \cite{Neerven}).

\section{Master double boxes}

The first master double box
\be
K(1,\ldots,1;d,s,t) =
\frac{\left(i \e^{-\gm_{\rm E}\ep} \right)^2 }{(-s)^{2+2\ep}(-t)}
K_1(t/s,\ep) \, ,
\label{K1def}
\ee
has been evaluated in ref.~\cite{K1} by use of $\alpha$-parameters
and resolving
singularities in a 5-fold MB integral:
\bea
K_1(x,\ep) &= &
-\frac{4}{\ep^4} +\frac{5\ln x}{\ep^3}
- \left( 2 \ln^2 x -\frac{5}{2} \pi^2  \right) \frac{1}{\ep^2}
\nn \\ && \hspace*{-30mm}
-\left( \frac{2}{3}\ln^3 x +\frac{11}{2}\pi^2 \ln x
-\frac{65}{3} \zeta(3) \right) \frac{1}{\ep}
+\frac{4}{3}\ln^4 x +6 \pi^2 \ln^2 x
-\frac{88}{3} \zeta(3)\ln x +\frac{29}{30}\pi^4 \nn \\ && \hspace*{-30mm}
- \left[
2 \Li{3}{ -x } -2\ln x \Li{2}{ -x }
-\left( \ln^2 x +\pi^2 \right) \ln(1+x)
\right] \frac{2}{\ep}
\nn \\ && \hspace*{-30mm}
- 4 \left(S_{2,2}(-x) - \ln x S_{1,2}(-x)  \right)
+ 44 \Li{4}{ -x } - 4 \left(\ln(1+x) + 6 \ln x  \right) \Li{3}{ -x }
\nn \\ && \hspace*{-30mm}
+ 2\left( \ln^2 x +2 \ln x \ln(1+x) +\frac{10}{3}\pi^2\right) \Li{2}{-x}
\nn \\ && \hspace*{-30mm}
+\left( \ln^2 x +\pi^2 \right) \ln^2(1+x)
-\frac{2}{3} \left(4\ln^3 x +5\pi^2 \ln x -6\zeta(3)\right) \ln(1+x) \; .
\label{K1}
\eea

To evaluate the second master double box, i.e.
\be
K(1,1,1,1,1,2,1;d;s,t) =
\frac{\left(i \e^{-\gm_{\rm E}\ep} \right)^2 }{(-s)^{2+2\ep} t^2}
K_2(t/s,\ep) \, ,
\label{K2def}
\ee
let us take first derivatives in $t$ of the two master double boxes.
Using $\alpha$-representation (\ref{KGenAlpha}) we obtain
\bea
\frac{\pa}{\pa t} K(1,\ldots,1;d;s,t) & =
& - K(1,1,1,1,2,2,2;d+2;s,t) \, ,
\label{RR1} \\
\frac{\pa}{\pa t}  K(1,1,1,1,1,2,1;d;s,t)& =
& - 2 K(1,1,1,1,2,3,2;d+2;s,t) \,.
\label{RR2}
\eea
We now use the results presented in Sect.~2 to express both right-hand sides
as linear combinations of the two master double boxes, vertex diagrams
(two in the first case and three in the second case) and a numerous
family (around fifty terms in each case) of diagonal crossed boxes and
boxes with one-loop insertions.
Substituting explicit result (\ref{K1}) into the first equation and
evaluating all the terms in the right hand side as explained in Section~3
we obtain an analytical result for the second master double box:
\bea
K_2(x,\ep) &= &
\frac{4}{\ep^4} - 5 \left(\ln x-2 \right)
\frac{1}{\ep^3}
+ \left( 2 \ln^2 x -14\ln x-\frac{5}{2} (\pi^2+4) \right) \frac{1}{\ep^2}
\nn \\ && \hspace*{-26mm}
+\left( \frac{2}{3}\ln^3 x+ 8\ln^2 x+\left(\frac{11}{2}\pi^2+14\right) \ln x
-2-3\pi^2-\frac{65}{3} \zeta(3) \right) \frac{1}{\ep}
\nn \\ && \hspace*{-26mm}
-\frac{4}{3}\ln^3 x (\ln x +1) -2\left(3 \pi^2+4\right) \ln^2 x
+ \left(10+9 \pi^2 +\frac{88}{3} \zeta(3) \right)\ln x
\nn \\ && \hspace*{-26mm}
+20+12\pi^2- \frac{29}{30}\pi^4 +\frac{4}{3}\zeta(3)
\nn \\ && \hspace*{-26mm}
+ x\left[
-\frac{7}{\ep^3} + \left(8\ln x-33 \right)
\frac{1}{\ep^2}
+ \left( 26 \ln x +6+\frac{21}{2}\pi^2 \right) \frac{1}{\ep}
\right.\nn \\ && \hspace*{-26mm} \left.
+\frac{1}{6} \left(
-32 \ln^3 x -4(21+26\pi^2)\ln x +180+209\pi^2+904\zeta(3)
\right)
\right]
\nn \\ && \hspace*{-26mm}
+ \left[
2 \Li{3}{ -x } -2\ln x \Li{2}{ -x }
-\left( \ln^2 x +\pi^2 \right) \ln(1+x)
\right] \frac{2}{\ep}
\nn \\ && \hspace*{-26mm}
-4 x \left[
8\left( \Li{3}{ -x }-\ln x \Li{2}{ -x }  \right)
-4\left( \ln^2 x +\pi^2 \right) \ln(1+x)
\right]
\nn \\ && \hspace*{-26mm}
+ 4 \left(S_{2,2}(-x) - \ln x S_{1,2}(-x)  \right)
- 44 \Li{4}{ -x } + 4 \left(\ln(1+x) + 6 \ln x  -2\right) \Li{3}{ -x }
\nn \\ && \hspace*{-26mm}
- 2\left( \ln^2 x +2 \ln x \ln(1+x)-4\ln x +\frac{10}{3}\pi^2\right) \Li{2}{-x}
-\left( \ln^2 x +\pi^2 \right) \ln^2(1+x)
\nn \\ && \hspace*{-26mm}
+\left(\frac{8}{3}\ln^3 x + 4\ln^2 x +\frac{10}{3}\pi^2 \ln x
+4\pi^2
-4\zeta(3)\right) \ln(1+x)
 \; .
\label{K2}
\eea

Proceeding in the same way with the second recurrence relation (\ref{RR2}),
and inserting there our analytical results for the master double boxes
we eventually obtain an identity of the left-hand and the right-hand sides.
This fact turns out to be a very non-trivial check of the recurrence
relations, their solutions and our analytical
expressions for the master double boxes.

\section{Asymptotic expansions of the double box}

We still want other checks and are going to compare our results
with what can be obtained by expanding the first master double box 
in various limits.
To expand the double box diagrams in the limit $t\to 0$ let us use
the strategy of regions:

({\it i})
Consider various regions of the loop momenta
and expand, in every region,
the integrand in a Taylor series with respect to the parameters
that are considered small in the given region;

({\it ii})
Integrate the integrand expanded, in every region in its own way,
over the whole integration domain in the loop momenta;

({\it iii})
Put to zero any scaleless integral.

In the off-shell and off-threshold limits,
this strategy leads to the well-known explicit prescriptions
\cite{offae} (see a brief review \cite{aerep}) based on the strategy
of subgraphs.
Although the strategy of subgraphs was successfully applied to
some on-shell limits \cite{onae,S2}, the strategy
of regions looks generally more flexible. In particular, it proved to be
adequate for constructing the threshold expansion \cite{BS}.

Let us choose, for convenience, the external momenta as follows:
\be
p_{1,2} = (\mp Q/2,Q/2,0,0), \;
r\equiv p_1+p_3 = (T/Q,0,\sqrt{T+T^2/Q^2},0) \, ,
\ee
where $s=-Q^2$ and $t=-T$.
The given limit is closely related to the Sudakov limit so that 
the following standard regions happen to be typical for it:
\bea
\label{h}
\mbox{{\em hard} (h):} && k\sim Q\, ,
\nn \\
\label{1c}
\mbox{{\em 1-collinear} (1c):} && k_+\sim Q,\,\,k_-\sim T/Q\, ,
\,\, \uk \sim \sqrt{T}\,,
\nn \\
\label{2c}
\mbox{{\em 2-collinear} (2c):} && k_-\sim Q,\,\,k_+\sim T/Q\, ,
\,\,\uk \sim \sqrt{T}
 \, , \nn \\
\label{us}
\mbox{{\em ultrasoft} (us):} && k\sim T\, .
\nn
\eea
Here $k_{\pm} =k_0\pm k_1, \; \uk=(k_2,k_3)$. We mean by $k\sim Q$, etc.
that any component of $k_{\mu}$ is of order $Q$.

It turns out that the (h-h), (1c-1c) and (2c-2c) are the only non-zero
contributions to the asymptotic expansion in the limit $t/s \to 0$.
In particular, all the (c-h) contributions and all the contributions with
ultrasoft momenta are zero because they generate scaleless integrals.

The (h-h) region generates the contribution given by Taylor expansion
of the integrand in the vector $r$. Every diagram from
this contribution corresponds
to the forward scattering configuration, $p_3=-p_1$ and $p_4=-p_2$, and can be
evaluated for general $\ep$ in gamma functions by resolving recurrence
relations following from integration by parts \cite{IBP}.
The first step of this procedure is to reduce an index $a_5$ or $a_7$ to zero
and thereby obtain vertex massless diagrams. The latter reduction, in the
scalar case, was constructed in ref.~\cite{S2}.
(In the case without numerators, the reduction of the forward scattering double
boxes was presented in ref.~\cite{DO}.)

We have constructed two different procedures for the evaluation
of the (h-h) part: along the lines
of this standard recursion and also by expanding the integrand of the
$\al$-representation in the variable $t$ and using tricks with shifting dimension.
We have implemented both methods and checked that they
give the same results for first several coefficients.

  We describe now how one can expand the integrals in the variable $t$.
The most suitable method to achieve this is the one proposed in \cite{OTqform}.
For the expansion in $t/s$, we have
\begin{eqnarray}
\label{ts}
  K^{(d)}(s,t) \z\stackrel{\rm h-h}{=}\z \sum_{j=0}^\infty
       \frac{1}{j!} \left(-\frac{t}{s}\right)^j
       Q_t^j K^{(d+2j)}(s,0)\,,
\end{eqnarray}
where $\,Q_t$ is the differential operator acting on the
masses of the lines.
If we denote $\partial/\partial m^2_i$
as $\partial_i$ then
\begin{eqnarray}
 Q_t \z=\z \partial_5\partial_6\partial_7\,.
\end{eqnarray}
After the differentiation in (\ref{ts}),
all these auxiliary masses are put to zero. As a result each coefficient in
the expansion (\ref{ts}) consists of integrals depending
only on one scale $s$.
These can already be evaluated analytically by the standard
``triangle'' rule, i.e. using the integration by parts \cite{IBP}.

  The integrals $K^{(d+2j)}(s,0)$ belong to the class of primitives,
i.e. they can be evaluated in terms of $\Gamma$-functions.
The problem here is that the repeated application
of the triangle rule brings more and more powers of
$1/\varepsilon$ and therefore a deeper expansion
of $\Gamma$-functions is required. To keep things
under control one can use e.g. an algorithm described below.
With the help of this algorithm, the depth of the $\varepsilon$-expansion is kept
at the level of 6. The reduction proceeds as follows:

  (i) Use the relation
\begin{eqnarray}\label{fkorobi}
\z\z ( 2d-2a_5-2a_6-2a_7
    -a_1 - a_2 - a_3 - a_4 )
        \nonumber\\
= \z\z a_1{\bf 1^+7^-}
   +a_2{\bf 2^+7^-}
   +a_3{\bf 3^+5^-}
   +a_4{\bf 4^+5^-}\,.
\end{eqnarray}
(This is the same first step as in the above mentioned standard recursive
procedure.)
With its help, we can get rid of either
line~5 or~7 and thereby reduce the double box to
a planar vertex. Relation~(\ref{fkorobi}) has the
feature that the left-hand side passes only once through
the ``critical point'' (when the expression in parentheses is proportional
to $\varepsilon$).
Therefore at most a single power of $1/\varepsilon$
will be generated in the course of this step of the recursion.

 (ii) Reduce indices of the lines 3 and 4 to one
with the help of
\begin{eqnarray}\label{fkorobii}
\z\z sa_3{\bf 3^+} =
   a_5{\bf 5^+4^-}+a_6{\bf 6^+}({\bf 4^-}-{\bf 2^-})
  +a_3{\bf 3^+4^-}
  -(d-2a_4-a_3-a_5-a_6)\,,\\
\z\z sa_4{\bf 4^+} =
   a_5{\bf 5^+3^-}+a_6{\bf 6^+}({\bf 3^-}-{\bf 1^-})
  +a_4{\bf 4^+3^-}
  -(d-2a_3-a_4-a_5-a_6)\,,
\end{eqnarray}
This step brings no new powers of $1/\varepsilon$.

 (iii) Shrink the line 4 (or 3) using the triangle rule
\begin{equation}
  (d-2a_6-a_1-a_2) =
   a_1{\bf 1^+}({\bf 6^-}-{\bf 3^-})
  +a_2{\bf 2^+}({\bf 6^-}-{\bf 4^-})\,.
\end{equation}
Here at most one additional power of $1/\varepsilon$ is generated.

 (iv) The index of the line 1 is then reduced to unity by
\begin{equation}
a_1{\bf 1^+} =
   (d-1-2a_7-a_6-a_1-a_2){\bf 7^+}
   +a_6{\bf 6^+7^+5^-}
   -a_2{\bf 2^+}
   -a_6{\bf 6^+}\,.
\end{equation}
No new powers of $1/\varepsilon$ arise here. Note that the line 7 could be absent
before this step but now it appears again
due to the $\bf 7^+$ term.

 (v) Apply the triangle rule
\begin{equation}
  (d-2a_3-a_5-a_6) =
   a_6{\bf 6^+}({\bf 3^-}-{\bf 1^-}) + a_5{\bf 5^+3^-} \,
\end{equation}
to shrink either line~1 or~3. Note that a dangerous
term $\bf 6^+1^-$ which leads to the oscillation
of the left-hand side around $\varepsilon$ is harmless here because
the index of the line~1 is already equal to unity due to (iv).

 (vi) Now we have to evaluate the diagram without lines~1 and~4.
(Other cases are trivial). To simplify it we
use (\ref{fkorobi}) which now looks like
\begin{eqnarray}\label{forwardkorob0}
 ( 2d-2a_5-2a_6-2a_7
    - a_2 - a_3 ) =
    a_2{\bf 2^+7^-}+a_3{\bf 3^+5^-}\,.
\end{eqnarray}
This completes the algorithm.

  Thus steps (i),(iii) and (vi) in the worst case
bring a factor $1/\ep$ each. Other three powers of $1/\ep$ arise
when evaluating primitive integrals. Therefore the total
depth of the $\varepsilon$-expansion of the $\Gamma$-functions
must be not greater than six.
  Note that diagrams without line~6 must
be separated from the very beginning and evaluated
immediately without any recursion in terms of $\Gamma$-functions. The point
is that when evaluating primitive diagrams with $a_6=0$ we obtain poles
of the fourth order instead of usual third order poles.

Let us now describe how an arbitrary term of the (c-c) contribution 
to the expansion of the (first) master double box can be
evaluated.
If we consider $k$ and $l$ 1-collinear then $k^2$, $l^2$, $p_2 k$ and $p_2 l$
are of order $T$ while $p_1 k$ and $p_1 l$ are of order $Q^2$. Moreover,
$(l+r)^2\equiv l^2+2lr-T \sim (l+\tilde{r})^2$, where $r=p_1+p_3$ and
\be
\tilde{r} = (T/(2Q),-T/(2Q),\sqrt{T},0) \, ,
\ee
with $2p_1 \tilde{r}=0, \; 2p_2 \tilde{r}=\tilde{r}^2=-T$.
Thus the (1c-1c) contribution is obtained by expanding propagators
$1/(k^2+2 p_1 k)$ and $1/(l^2+2 p_1 l)$ in Taylor series respectively
in $k^2$ and $l^2$,
and by expansion also in Taylor series in $2p_1 r$.
(Note that we are dealing with a function of three kinematical
variables, $2p_1 r, \, 2p_2 r$ and $r^2$. So, we expand the integrand
(e.g. in the $\alpha$-representation) in $2p_1 r$ and then put $2p_1 r=T$.)
Actually we want to expand the master box with all $a_i$ equal to $1$.
Therefore only the leading term in the Taylor expansion 
in $k^2$ is non-zero because, starting from the next order, 
the factor $k^2$ cancels the
propagator $1/k^2$ and we obtain a zero scaleless integral.

In the following it is more convenient to consider
integrals with general arbitrary indicies $a_i$.
Thus, we need integrals of the type
\bea
J(a_1,\ldots,a_9;d,s,t) &=&
\frac{1}{a_8!} \left(\frac{\pa}{\pa X}\right)^{a_8}
\int\int \frac{\dd^dk \dd^dl}{
(\pi^{d/2})^2 (-2 p_1 l)^{a_1} (-l^2+2 p_2 l)^{a_2}} \nn \\ && \hspace*{-30mm}
\times
\left. \frac{(l^2)^{a_9}}{
(-2 p_1 k)^{a_3} (-k^2+2 p_2 k)^{a_4}
(-k^2)^{a_5} (-(k-l)^2)^{a_6} (-(l-r)^2)^{a_7}
} \right|_{X=0}\,,
\label{2boxGen}
\eea
with $X=2p_1r$, for integer $a_i$.
However this integral taken alone is not generally regularized dimensionally. Only
if we add the corresponding symmetrical contribution (i.e.
for $a_1 \leftrightarrow a_2, \, a_3 \leftrightarrow a_4$)
we shall have
a result that exists within dimensional regularization. An
efficient way to deal with this problem is to introduce
an auxiliary analytical regularization which allows to consider the above
terms separately. Let us introduce it
into the lines 1 and 2 (although we could choose as well 3 and 4, or all
four of these lines), i.e. with $a_1\to a_1+x_1,
a_2 \to a_2+x_2$ plus a symmetrical contribution
which is given by interchanging
$a_1+x_1$ and $a_2+x_2$. Only in the sum we may switch off this
regularization, i.e. let $x_1\to 0, x_2\to 0$.

The general integral (\ref{2boxGen}) at $a_8=a_9=0$ can be represented as
\bea
J(a_1,\ldots,a_7,0,0;d,s,t) &=&
\frac{i^2 (-1)^a \Gm(a'-d)}{
\prod_{i\neq 1,3} \Gm(a_i) (Q^2)^{a_1+a_3} T^{a'-d}} \nn \\&& \hspace*{-33mm}
\times \int \dd \ual' \dl\left( \sum_{i\neq 1,3} \al_i-1\right)
\prod_{i\neq 1,3} \al_i^{a_i-1}
(\al_5 \al_6\al_7)^{d-a'} \bar{D}^{a-\frac{3}{2}d} D_1^{-a_1}
D_3^{-a_3} \, ,
\label{2boxParam}
\eea
where
\bea
\bar{D}&=&(\al_2+\al_7) (\al_4+\al_5) +\al_6 (\al_2+\al_4+\al_5+\al_7) \,, \\
D_1&=&\al_2 (\al_4+\al_5) + \al_6 (\al_2+\al_4) \,, \\
D_3&=&\al_4 (\al_2+\al_7) + \al_6 (\al_2+\al_4) \,. \\
\eea
Moreover, the integral above is represented in terms of parameters
$\ual'=\{ \al_i\,, i=2,4,5,6,7\} \,,
a =\sum a_i \,$,\linebreak
 $a' =\sum_{i\neq 1,3} a_i \,$.
In the argument of the delta function one can put
the sum of an arbitrary subset of $\al_i\,, i=2,4,5,6,7$.

The integrals with $a_8>0$ are obtained by the replacement
\[
(Q^2)^{-a_1-a_3} D_1^{-a_1} D_3^{-a_3} \to
(Q^2 D_1+ X B_1 )^{-a_1} (Q^2 D_3+ X B_3 )^{-a_3}\, ,
\]
where $B_1=\al_7 (\al_4+\al_5+\al_6)$ and $B_3=\al_6 \al_7$,
subsequent expansion in Taylor series in $X$ and keeping terms of order
$X^{a_8}$.

Starting from (\ref{2boxParam}) and it generalization for $a_8>0$
we can arrive at a triple MB integral in the following way:

1. Choose the delta function as $\dl(\al_4+\al_6-1)$.
2. Represent $1/(D_3)^{a_3+\ldots}$ in a MB representation,
using the decomposition $D_3=D_0+\al_4 \al_7$.
3. Integrate in $\al_7$. Note that the function $D_1+ \al_5 \al_6$
arises.
4. Represent  $(D_1+ \al_5 \al_6)^{\ldots}$ via MB representation.
5. Represent  $(D_1=D_0+ \al_2 \al_5)^{\ldots}$ via MB
representation.
6. Integrate in $\al_5$ and then in $\al_2$.
7. Integrate in $\xi=\al_4$ with $\al_6=1-\xi$.

Then we obtain a triple MB integral of a ratio of $\Gamma$-functions.
This integral is evaluated, in expansion in $\ep$, by the standard technique
of shifting contours and expansion in MB integrals.
First, singularities in $x_1-x_2$ are localized.
Second, the same technique is applied for picking up singularities in
$\ep$.
As a result we end up with a collection of explicit terms plus
integrals which are finite in $\eps$. These last integrals can be expanded
in $\eps$ up to the desired order. So, in the end, various
integrals with $\Gamma$, $\psi$ and  their derivatives
are evaluated (see examples in \cite{K1}).

The integrals with $a_9>0$ can be represented in $\al$-parameters in a
cumbersome way. Using IBP \cite{IBP} (starting from the integral of
$\left( \pa/{\pa k}\right) \,k $) we obtain the following recurrence 
relation:
\be
\left(d-a_3-a_4- 2 a_5- a_6-a_4 \pppp \mmmmm
-a_6 \pppppp (\mmmmm -\pnine) \right) J =0 \, .
\label{RR-cc}
\ee
Note that our general integral is zero when at least one of parameters
$a_5, a_6, a_7$ is a non-positive integer so that this relation
is of no use in its primary form. However we may turn to a new integral
which is defined without $\Gm(a_6)$ in the denominator, i.e. put
$J(\ldots a6,\ldots) = J'(\ldots a6,\ldots)/\Gm(a_6)$.
Then (\ref{RR-cc}) applied to the new integrals $J'$ 
can be used to decrease the number $a_9$ to zero.
We shall generally deal with non-positive $a_6$ after that.

To evaluate the (c-c) contribution to the expansion of the
first master double box it suffices to consider
$a_2=a_3=a_4=a_5=a_7=1$, and general integer $a_1, a_6$ and
$a_8,a_9\geq 0$. Then $a_6$ can be reduced to zero using
(\ref{RR-cc}). In particular, in the leading order of the expansion we meet
$J(1,\ldots,1,0,0)$ and, in the NLO, $J(2,1\ldots,1,0,1)$ and
$J(1,1\ldots,1,1,0)$. Using (\ref{RR-cc}) we then express $J(2,1\ldots,1,0,1)$
through $J(2,1,1,1,1,0,1,0,0)$ (defined at negative $a_6$ as explained above).

We have evaluated the LO and NLO of the (c-c) contribution
(and added the LO order of the (h-h) contribution which is really
NLO for the whole expansion)
of the master double box integral in the limit $t/s \to 0$.
After that we have found complete
agreement with the first two terms of the expansion of
the explicit result (\ref{K1}):
\bea
K_1(x,\ep) &=&
-\frac{4}{\ep^4} +\frac{5\ln x}{\ep^3}
- \left( 2 \ln^2 x -\frac{5}{2} \pi^2  \right) \frac{1}{\ep^2}
\nn \\ &&  \hspace*{-20mm}
-\left( \frac{2}{3}\ln^3 x +\frac{11}{2}\pi^2 \ln x
-\frac{65}{3} \zeta(3) \right) \frac{1}{\ep}
\nn \\ && \hspace*{-20mm}
+\frac{4}{3}\ln^4 x +6 \pi^2 \ln^2 x
-\frac{88}{3} \zeta(3)\ln x +\frac{29}{30}\pi^4
\nn \\ && \hspace*{-20mm}
+ 2 x \left[
\frac{1}{\ep}\left(\ln^2 x - 2 \ln x +\pi^2+2\right)
\right.\nn \\ && \hspace*{-20mm}
\left.
-\frac{1}{3} \left(
4\ln^3 x +3 \ln^2 x + (5\pi^2-36)\ln x +2 (33+5\pi^2-3\zeta(3))
\right)
\right] 
\nn \\ && \hspace*{-20mm}
 + O(x^2 \ln^3 x) \, .
\label{NLO}
\eea

The expansion in the limit $s/t\to 0$ has the same structure as in the previous
case: only (h-h) and (c-c) contributions are non-zero. However, in this case,
there are three (c-c) contributions, and the poles with respect to
an auxiliary parameter of analytic regularization happen
to be up to the second order (as in the case of the non-planar
vertex diagram in the Sudakov limit --- see \cite{SR}).

\section{Conclusion}

The on-shell double box provides a curious example
where the analytical evaluation is simpler than the evaluation
by means of asymptotic expansions in some limits (in this case,
$t/s\to 0$ and $s/t\to 0$). We have met rather inconvenient recurrence
relations for diagrams that enter the collinear-collinear contribution.
Even the global recurrence relations for the unexpanded diagram
turned out to be simpler, with the solutions described in Section~2.
Although a systematical evaluation of the (c-c) contribution ot an arbitrary
order in the expansion is certainly possible,
it would be difficult to guess the analytic form ot the result
by evaluating and studying first terms of the
expansion. Still we really used, in two points,
the method of asymptotic expansions for
crucial checks of our explicit procedure. Firstly, we have used the possibility
to evaluate an arbitrary term of the (h-h) contribution (both in the
limits ($t/s\to 0$ and $s/t\to 0$) for checking (global) recurrence relations
and their solutions presented in Section~2. This was possible because
any recurrence relation derived within integration by parts commutes with
the (h-h) expansion which, by definition, is performed under the sign
of the integrals involved (either in the integrals in the loop momenta or
in the $\alpha$-parameters).
Secondly, we have evaluated
the LO and NLO contributions of the first master double box and
successfully compared them with the expansion of the analytical result.

In fact, we could avoid the rather
cumbersome evaluation of the first master double
box by starting from equations (\ref{RR1}) and (\ref{RR2}),
applying solutions of recurrence relations of Section~2,
evaluating all simpler diagrams, eliminating
the second master double box by use of the second equation and arriving
at a second order differential equation for the first master double box.
It would be possible to solve this equation by expanding in $t/s$.
But then we would need boundary conditions to solve it.
Here we could insert the first two orders of the expansion in the limit
$t/s\to 0$ evaluated by means of the strategy of regions as explained in
Section~5 (where we have met only 3-fold rather 5-fold MB integrals.)
However, in this case, we would not have crucial checks for the obtained results.

\vspace{0.5 cm}

{\em Acknowledgments.}
We want to thank L.J.~Dixon for finding several misprints in the text
and communication concerning cross-checks of the recurence relations.
The work of V.S. was supported by the Volkswagen Foundation, contract
No.~I/73611, and by the Russian Foundation for Basic Research,
project 98--02--16981.


\begin{thebibliography}{99}
\bibitem{dimreg}
G.~'t Hooft and M.~Veltman, {\em Nucl.~Phys.} B44 (1972) 189;
C.G.~Bollini and J.J.~Giambiagi, {\em Nuovo Cim.} 12B (1972) 20.


\bibitem{K1}
V.A.~Smirnov, {\em Phys. Lett.} B460 (1999) 397.


\bibitem{V1}
O.L. Veretin, to be published.

\bibitem{IBP}
K.G.~Chetyrkin and F.V.~Tkachov, {\em Nucl. Phys.} B192 (1981) 159.

\bibitem{Breitenlohner}
P. Breitenlohner and D. Maison, {\em Commun. Math. Phys.} 52 (1977) 39.

\bibitem{OTdim}
O.V. Tarasov, {\em Phys. Rev.} D54 (1996) 6479.

\bibitem{Lewin}
L.~Lewin, {\em Polylogarithms and associated functions}
      (North Holland, 1981).

\bibitem{GenPolyLog}
K.S. Kolbig, J.A. Mignaco and E. Remiddi, {\em B.I.T} 10 (1970) 38;
K.S. Kolbig, Math. Comp. 39 (1982) 647;
A.~Devoto and D.W.~ Duke, {\em Riv. Nuovo Cim.} 7 (1984) 1.

\bibitem{Oleg}
J. Fleischer, A.V. Kotikov and O.L. Veretin,
{\em Nucl. Phys.} B547 (1999) 343.

\bibitem{Neerven}
W.L. van Neerven, {\em Nucl.Phys.} B268 (1986) 453.

\bibitem{offae}
S.G.~Gorishny, preprints JINR E2--86--176, E2--86--177 (Dubna 1986);
{\em Nucl. Phys.} B319 (1989) 633;
K.G.~Chetyrkin, {\em Teor. Mat. Fiz.} 75 (1988) 26; 76 (1988) 207;
K.G.~Chetyrkin, preprint MPI-PAE/PTh 13/91 (Munich, 1991);
V.A.~Smirnov, {\em Commun. Math. Phys.} 134 (1990) 109;
V.A.~Smirnov, {\em Renormalization and asymptotic expansions}
(Birkh\"{a}user, Basel, 1991).

\bibitem{aerep}
V.A.~Smirnov, {\em Mod. Phys. Lett.} A 10 (1995) 1485.

\bibitem{onae}
V.A.~Smirnov, {\em Phys. Lett.} B394 (1997) 205;
A.~Czarnecki and V.A.~Smirnov, {\em Phys. Lett.} B394 (1997) 211.

\bibitem{S2}
V.A.~Smirnov, {\em Phys. Lett.} B404 (1997) 101.

\bibitem{BS}
M. Beneke and V.A.~Smirnov, {\em Nucl. Phys.} B522 (1998) 321.

\bibitem{DO}
A.I. Davydychev and P. Osland, {\em Phys. Rev.} D59 (1998) 014006.

\bibitem{OTqform}
O.V. Tarasov, {\em Nucl. Phys.} B480 (1996) 397.

\bibitem{SR}
V.A.~Smirnov and E.R. Rakhmetov, {\em Teor. Mat. Fiz.} 120 (1999) 64.

\end{thebibliography}
\end{document}